\documentclass[journal,draftcls,onecolumn,12pt,twoside]{IEEEtran}

\usepackage{moreverb}
\usepackage{hhline}
\usepackage{amssymb,amsfonts,amsthm}
\usepackage{amsmath}
\usepackage{amsfonts}
\usepackage{epsfig}
\usepackage{color}
\usepackage{enumerate}
\usepackage{ifpdf}
\usepackage{graphicx}
\usepackage{subfig}
\usepackage{caption}
\usepackage[font=footnotesize,skip=0pt]{caption}
\usepackage{multirow}
\usepackage{mathtools}
\usepackage{algorithm,algpseudocode}
\usepackage{stfloats}
\usepackage{csquotes}

\newcommand\floor[1]{\lfloor#1\rfloor}
\newcommand\ceil[1]{\lceil#1\rceil}

\begin{document}

\title{Finite-length performance comparison of network codes using random vs Pascal matrices}

\author{Tan Do-Duy
        and~M. \'{A}ngeles V\'azquez-Castro
\thanks{Tan Do-Duy is with the Department of Computer and Communication Engineering, HCMC University of Technology and Education, Vietnam, e-mail: tandd@hcmute.edu.vn.}
\thanks{M. \'{A}ngeles V\'azquez-Castro is with the Department of Telecommunications and Systems Engineering, School of Engineering, Autonomous University of Barcelona, Spain, e-mail: angeles.vazquez@uab.es.}
}

\maketitle

\begin{abstract}
In this letter, we evaluate the finite-length performance of network coding when using either random or structured encoding matrices. 
First, we present our novel construction of structured network codes over $\mathbb{F}_q$ $(q=2^m)$ using Pascal matrices. We present their encoding, re-encoding and decoding in matrix notation and derive their packet loss rate. Second, we propose a novel methodology to compute the optimal finite-length coding rate for representative and realistic traffic applications. 
Finally, our method allows to compare the performance of our codes with the performance of popular random codes.
We show that our constructions always have better throughput and minimal overhead, which is more significant for short code lengths. Further, their larger decoding delay fulfils the delay constraints of realistic scenarios (e.g. 5G multihop networks).
\end{abstract}


\section{Introduction \label{Sec:INTRODUCTION}}
Linear network coding \cite{Li.2003, SARSHAR.2014, Yang.2019} has been studied as a practical network coding (NC) scheme for improving throughput and reliability for wireless networks. It can be roughly classified into structured linear NC and random linear NC, depending on whether the encoding matrices are chosen either randomly or with (algebraic) structure. Structured codes need decoding and re-encoding at in-network nodes thus increasing the delay. On the other hand, random codes need to solve the signaling problem of the randomly-chosen coefficients \cite{Saxena.2015}. 

In this letter, we consider a simple, but common in practice, network where a source sends a sequence of packets to a destination over a line network of erasure links e.g. relay-aided networks \cite{Saxena.2015, Baofeng.2018, Baofeng.2019}. For each NC scheme, we model the encoding, re-encoding and decoding process in matrix notation and derive the theoretical expression of packet loss rate (PLR) at the destination. Then, we propose a methodology to compute the optimal coding rate to fulfil application's reliability targets for any erasure rates and number of links. We show that our constructions of structured codes over $\mathbb{F}_q$ using Pascal matrices always have better throughput and minimal overhead. Even if the inventors of the Pascal matrices we use \cite{Hua.2018} do mention their applicability to NC, we are the first ones to propose practical implementation and performance analysis.
Moreover, existing studies on practical applications of network codes usually assume a fixed size of information packets \cite{Shrader.2009} or capacity achievability \cite{Saxena.2015} where the coding block size is assumed to tend to infinity, which is not a practical assumption.
We thus study the performance of different NC schemes with finite-length coding rate.

\section{System Model \label{Sec:SYSTEM_MODEL}}
A line network connecting a source-destination pair is modeled as an acyclic graph $\mathcal{G}$ with $\mathcal{N}$ nodes and $\mathcal{L}$ links. $\mathcal{D}$ is the set of random variables corresponding to the erasure process associated with each link. We consider random packet erasure model where each link $i$ is modeled as a memoryless channel with erasure probability $\delta_{i}$ $(1 \leq i \leq \mathcal{L})$. For a line network, the theoretical capacity is $\underset{1\le i\le \mathcal{L}}{\text{min}} \hspace{0.5mm} \big(1-\delta_i \big)$. Here, we assume all the erasure processes are equal $(\mathcal{D}=\delta)$. 
A packet stream is produced at the source. For a finite field $\mathbb{F}_q$ $(q=2^m)$, we use $M$ symbols as the packet length. 
We call \enquote{generation} a group of $K$ information packets, denoted as $X \in \mathbb{F}^{M\times K}_q$. We assume per-generation block coding with block-length $N$ and $\rho=\frac{K}{N}$ is the coding rate. 
$P_e(\delta,\rho)$ is the PLR at the destination after decoding. 
We define achievable rate at the destination as $R = \rho \big(1 - P_e(\delta,\rho)\big)$. 
In this letter, we propose a methodology to compute the optimal coding rate, $\rho^*$, that provides the optimal achievable rate, $R^*$, that meets a given PLR target, $P_e^0$.
Coding overhead ($\%$) is defined as $\gamma=\frac{100\sigma}{M}$ with $\sigma$ the amount of symbols needed in a packet to signal coding coefficients.

\section{Linear Network Coding Schemes \label{Sec:NETWORK_CODES}}
\begin{figure}[htbp]
\begin{centering}
\subfloat[PascalNC				\label{fig:ILLUSTRATION_SRNC}]				{\epsfig{file=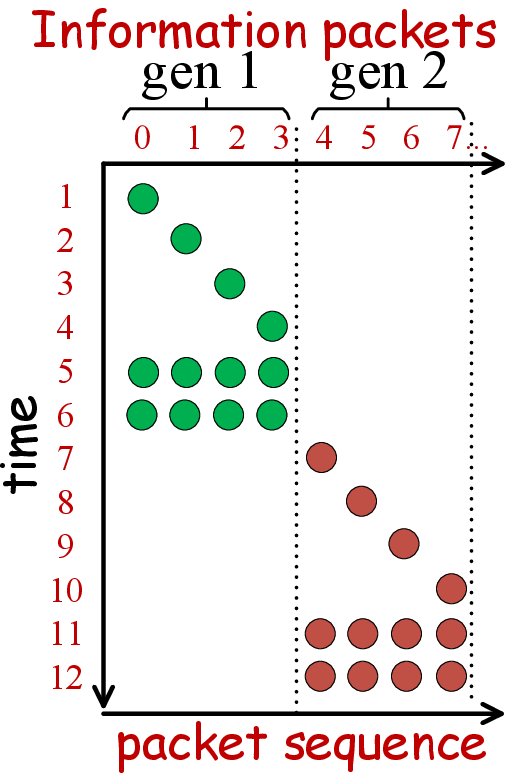,width=0.206\textwidth}}
\hspace{8 mm}
\subfloat[PascalNC-S			\label{fig:ILLUSTRATION_PACE}]				{\epsfig{file=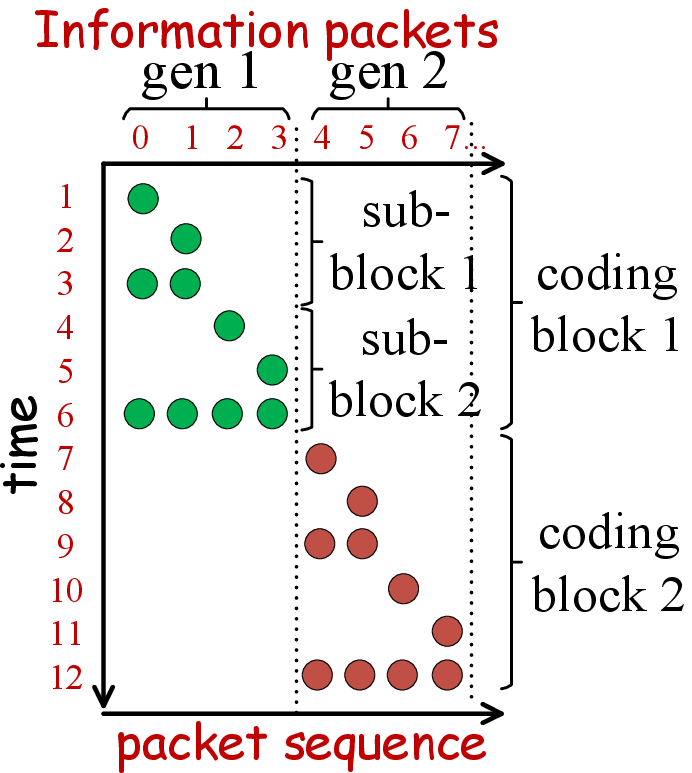,width=0.28\textwidth}}
\par\end{centering}
\caption{Illustration of PascalNC and PascalNC-S for $K=4$, $N=6$. Each single dot represents an uncoded packet while multiple dots on a line represent a coded packet.
\label{fig:ILLUSTRATION}}
\end{figure}

\subsection{Structured Linear NC using Pascal matrices}
While any maximum distance separable matrix of algebraic codes can be used for NC, we have chosen Pascal matrices as proposed in \cite{Hua.2018} due to their property of easy regeneration of coefficients at the nodes. The Pascal matrix  $P_{q}$ of size $q\times q$ is the traditional Pascal matrix $modulo$-$q$. To encode $K$ information packets, the first $K$ symbols of Pascal matrix columns form the columns of the coding coefficient matrix. The choice of columns can be random or optimized offline, which may or may not contain zeros.  Intermediate nodes decode and re-encode before forwarding packets and decoding is with progressive Gauss-Jordan elimination \cite{Saxena.2015} according to signaled columns or stored tables. In case of offline optimization, different solutions are being proposed for software updates of dynamically created nodes (e.g. see \cite{MAngeles2.2018}). 

\subsubsection{Systematic Pascal NC (PascalNC)}
Let $X'=XG$ represent $N$ packets transmitted by the encoder (as illustrated in Fig. \ref{fig:ILLUSTRATION_SRNC}).
The generator matrix is $G=\left[\begin{matrix} I_{K} ~ C \end{matrix}\right]$. It consists of the identity matrix $I_{K}$ of dimension $K$ and a coefficient matrix $C\in\mathbb{F}_{q}^{K \times (N-K)}$ extracted from the first $K$ symbols of $N-K$ columns of the Pascal matrix $P_{q}$.

We denote a binomial r.v. $V \sim bin(n,p)$ with the probability mass function $Pr(V=v)=\binom{n}{v} p^{v}\left(1-p \right)^{n-v}$, $v=0,1,...,n$ with $p=1-\delta$. For space limitation, here we assume Pascal codes with deterministically chosen columns. Let $Z_i^{(ps)}$ be the random number of packets decoded at a node $i$.
$Pr(Z_i^{(ps)}=x<K)$ and $Pr(Z_i^{(ps)}=K)$ can be obtained by modifying Eqs. (7)-(8) of \cite{Shrader.2009}.

The PLR for PascalNC over a single-hop $i$ $(1\le i \le \mathcal{L})$ is

\begingroup\makeatletter\def\f@size{10}\check@mathfonts
\begin{align}
 P_{e_i}^{(ps)}(\delta,\rho) & = 1 - \frac{1}{K} \sum_{z=1}^{K}zPr(Z_i^{(ps)} = z).	\nonumber
\end{align}
\endgroup

Therefore, the PLR for PascalNC at the destination in $\mathcal{G}$ is bounded as

\begingroup\makeatletter\def\f@size{10}\check@mathfonts
\begin{align}
P_e^{(ps)} \le 1 - \prod_{i=1}^{\mathcal{L}}\Big(1 - P_{e_i}^{(ps)}(\delta,\rho)\Big).
\label{eq:PE_PascalNC}
\end{align}
\endgroup

\subsubsection{Systematic Pascal NC with Scheduling (PascalNC-S)}
For PascalNC-S, we consider two sub-blocks as illustrated in Fig. \ref{fig:ILLUSTRATION_PACE}. The first $K_1=\lceil \frac{K}{2} \rceil$ information packets are transmitted first and followed by $n_{c_1}=\lceil \frac{N-K}{2} \rceil$ coded packets. Then, the last $K_2=K-K_1$ information packets and $n_{c_2}= N-K-n_{c_1}$ coded packets combining $K$ information packets are transmitted in sequence. 
Hence, the SNC-S generator matrix is $G = \big[I_1 ~ C_1 ~ I_2 ~ C_2\big]$, where

\begin{itemize}
	\item $I_1$: an $K\times K_1$ matrix, the first $K_1$ columns of the identity matrix $I_{K}$.
	\item $C_1$: an $K\times n_{c_1}$ matrix where elements of the first $K_1$ rows are extracted from the first $K_1$ symbols of $n_{c_1}$ columns of the Pascal matrix  while the remaining elements are zero.
	\item $I_2$: an $K\times K_2$ matrix, the last $K_2$ columns of the identity matrix $I_{K}$.	
	\item $C_2$: an $K\times n_{c_2}$ matrix where elements are extracted from the first $K$ symbols of $n_{c_2}$ columns of the Pascal matrix.
\end{itemize}

For a single hop $i$, we denote $U_s \sim bin(K_s,p)$ and $E_s \sim bin(n_{c_s},p)$ as the number of uncoded and coded packets of the $s$-th sub-block $(s=1,2)$ received at the decoder $i$, respectively. The decoder can recover the first $K_1$ information packets whenever it receives enough $K_1$ packets of the first sub-block.
Let $Z_{i_{1}}^{(pc)}$ and $Z_{i}^{(pc)}$ be the number of packets of the first $K_1$ information packets and the total number of packets of the generation decoded, respectively. 
$Z_{i}^{(pc)}<K_1$ indicates that the decoder is only able to decode some uncoded packets of the first and the second sub-block. $K_1 \le Z_{i}^{(pc)}<K$ indicates that the decoder can decode all or some uncoded packets of the first sub-block and is only able to decode some uncoded packets of the second sub-block. 
For $x=1,2,..., K_1-1$,

\begingroup\makeatletter\def\f@size{10}\check@mathfonts
\begin{align}
Pr(Z_{i}^{(pc)}=x) & =\sum_{u_1=0}^{x}	Pr(Z_{i_{1}}^{(pc)}=u_1)	Pr(Z_{i}^{(pc)}=x|Z_{i_{1}}^{(pc)}=u_1),
\label{eq:PascalNC_S1}
\end{align}
\endgroup

and for $x=K_1, K_1+1,..., K-1$,		

\begingroup\makeatletter\def\f@size{10}\check@mathfonts
\begin{align}						
Pr(Z_{i}^{(pc)}=x)	&	=	\sum_{u_1=0}^{K_1-1}Pr(Z_{i_{1}}^{(pc)}=u_1)Pr(Z_{i}^{(pc)}=x|Z_{i_{1}}^{(pc)}=u_1)	+ Pr(Z_{i_{1}}^{(pc)}=K_1)	Pr(Z_{i}^{(pc)}=x|Z_{i_{1}}^{(pc)}=K_1),	\\				
Pr(Z_{i}^{(pc)}=K)	&	=	\sum_{u_1=0}^{K_1-1}	Pr(Z_{i_{1}}^{(pc)}=u_1)	Pr(Z_{i}^{(pc)}=K|Z_{i_{1}}^{(pc)}=u_1)	+ Pr(Z_{i_{1}}^{(pc)}=K_1)	Pr(Z_{i}^{(pc)}=K|Z_{i_{1}}^{(pc)}=K_1),
\label{eq:PascalNC_S3}
\end{align}

\begin{align}						
with ~ Pr(Z_{i_{1}}^{(pc)}=u_1<K_1)		& = Pr(U_1=u_1)	\Big(1 - Pr\big(E_1\geq K_1-u_1\big)\Big)	\hspace{15mm}		\nonumber	\\
													&	= Pr(U_1=u_1)	\big(1-\sum_{e=K_1-u_1}^{n_{c_1}}	Pr(E_1=e)\big),			\nonumber
\end{align}

\begin{align}						
Pr(Z_{i}^{(pc)}=x<K|Z_{i_{1}}^{(pc)}=u_1<K_1)	= Pr(U_2=x-u_1) (1-Pr(E_1< K_1-u_1,E_2\geq K-E_1-x) )				\nonumber	\\
= Pr(U_2=x-u_1)	(1-\sum_{e_1=0}^{K_1-u_1-1}Pr(E_1=e_1)\sum_{e=K-e_1-x}^{n_{c_2}}Pr(E_2=e))	\nonumber
\end{align}

and by similar arguments, we can write

\begin{align}						
Pr(Z_{i_{1}}^{(pc)}=K_1)					&	= \sum_{u_1=0}^{K_1}Pr(U_1=u_1)	\sum_{e=K_1-u_1}^{n_{c_1}}Pr(E_1=e),	\hspace{5mm}\nonumber	\\
Pr(Z_{i}^{(pc)}=x<K|Z_{i_{1}}^{(pc)}=K_1)		&	= Pr(U_2=x-K_1)	(1-\sum_{e=K-x}^{n_{c_2}}Pr(E_2=e))				\nonumber
\end{align}

\begin{align}						
Pr(Z_{i}^{(pc)}=K|Z_{i_{1}}^{(pc)}=u_1<K_1)	= \sum_{u_2=0}^{K_2}Pr(U_2=u_2)	\sum_{e_1=0}^{K_1-u_1-1}	Pr(E_1=e_1)	\sum_{e=K-u_1-e_1-u_2}^{n_{c_2}}Pr(E_2=e),	\nonumber
\end{align}

\begin{align}						
Pr(Z_{i}^{(pc)}=K|Z_{i_{1}}^{(pc)}=K_1)	 		=\sum_{u_2=0}^{K_2}Pr(U_2=u_2)	\sum_{e=K_2-u_2}^{n_{c_2}}	Pr(E_2=e).	\nonumber
\end{align}
\endgroup

Hence, the PLR for PascalNC-S over a single-hop $i$ is

\begingroup\makeatletter\def\f@size{10}\check@mathfonts
\begin{align}
P_{e_i}^{(pc)}(\delta,\rho) = 1 - \frac{1}{K} \sum_{z=1}^{K}zPr(Z_i^{(pc)}=z).		\nonumber
\end{align}
\endgroup

Therefore, the PLR for PascalNC-S at the destination is bounded as

\begingroup\makeatletter\def\f@size{10}\check@mathfonts
\begin{align}
P_e^{(pc)} \le 1 - \prod_{i=1}^{\mathcal{L}}\Big(1 - P_{e_i}^{(pc)}(\delta,\rho)\Big).
\label{eq:PE_PascalNC_S}
\end{align}
\endgroup

\subsection{Random Linear NC}
We consider systematic random linear capacity-achieving and sub-optimal codes: Systematic Random NC (SNC) and Systematic Random NC with Scheduling (SNC-S). 
Coding coefficients are chosen randomly from the same $\mathbb{F}_q$. 
To signal coding coefficients, $\mathcal{L}$ symbols in each packet are invested in sending random seeds while $K$ symbols are invested in sending coding coefficients. 
The PLR for SNC is derived in \cite{Shrader.2009} and the PLR for SNC-S can be deduced from Eqs. (\ref{eq:PascalNC_S1})-(\ref{eq:PascalNC_S3}).

From the perspective of linear coding operation \cite{Li.2003}, the encoding and decoding complexity of the codes is $\mathcal{O}(MK)$ and $\mathcal{O}(K^3+MK^2)$, respectively. The complexity is kept low due to the high sparsity property of the generator matrix.

\section{Network Coding Rate Optimization \label{Sec:OPTIMIZATION}}
Given $N$ and $P_e^0$, the optimal coding rate, $\rho^*$, is defined as

\begingroup\makeatletter\def\f@size{10}\check@mathfonts
\begin{equation} 
\begin{aligned} 
& & & \rho^* = \underset{\rho \in \Psi}{\text{max}} \hspace{1mm} \rho \\
& \text{s.t.}
& & 	P_e(\delta,\rho) 		\le		P_e^{0},
\end{aligned} 
\label{Eq:OPTIMIZATION}
\end{equation}
\endgroup

where $\Psi$ is the set of coding rates available for searching, $\rho[i]=i\rho_0$, $1\leq i\leq |\Psi|$ (e.g., $\rho_0=\frac{1}{N}$, $|\Psi|=N-1$). The PLR at the destination, $P_e(\delta,\rho)$, is obtained from Eq. (\ref{eq:PE_PascalNC}) and Eq. (\ref{eq:PE_PascalNC_S}) for the case of PascalNC and PascalNC-S, respectively.

It is known that the PLR is a nondecreasing function with $\rho$ \cite{Shrader.2009}. Thus, to solve Problem (\ref{Eq:OPTIMIZATION}), we propose Algorithm \ref{Algo:SEARCHING} based on the binary searching algorithm \cite{Trenchea.2014}.
Algorithm \ref{Algo:SEARCHING} converges in $k' = \ceil{log_2(|\Psi|-1)}$ iterations. Due to space limitations we do not give the detailed proof.

\begin{algorithm}
\caption{The proposed searching algorithm to identify $\rho^*$.}
\label{Algo:SEARCHING}
\begin{algorithmic}[1]
\State 			\textbf{Initialize}
\State \hspace{5pt}	Set $\mathcal{L}$, $\delta$, $m$, $N$, $P_e^0$, $\Psi$, $first=1$, $last=|\Psi|$, $\rho^*=\emptyset$
\State 			\textbf{while}	$(first<last)$	\hspace{4mm}	\{	\hspace{2mm}	$mid=\floor{\frac{first+last}{2}}$
\State \hspace{8pt}		Compute $P_e(\delta,\rho[mid])$	(theory or simulations)
\State \hspace{8pt}  \textbf{if}			 $mid=first$	\hspace{2mm}	\textbf{then}		
\State \hspace{20pt}							\textbf{if} 	$P_e(\delta,\rho[mid])\le P_e^0$		\hspace{1mm}	\textbf{then}		\hspace{1mm}	$\rho^{*}=\rho[mid]$	\hspace{1.5mm}	\textbf{Break}
\State \hspace{8pt}	\textbf{else~if} $P_e(\delta,\rho[mid])\le P_e^0$							\textbf{then}	$first=mid$, $\rho^{*}=\rho[mid]$
\State \hspace{8pt}  \textbf{else}		\hspace{1mm}		$last=mid$	\hspace{2mm}	\}
\State    	Return $\rho^*$
\end{algorithmic}
\end{algorithm}

\section{Performance Evaluation \label{Sec:PERFORMANCE}}
This section evaluates the performance of NC schemes with $P_e^0=10^{-6}$ and $P_e^0=10^{-3}$ representing quasi-error-free channel and delay-constrained applications \cite{Chen.2004}, respectively. We consider $m=8$, $M=1500$, $\delta=0.05$ and $\delta=0.2$ representing 802.11 wireless links \cite{Salyers.2008} and space scenarios with links impacted by light rainfall \cite{Saxena.2015}, respectively.

\begin{figure*}[htbp]
\begin{center}
\epsfig{file=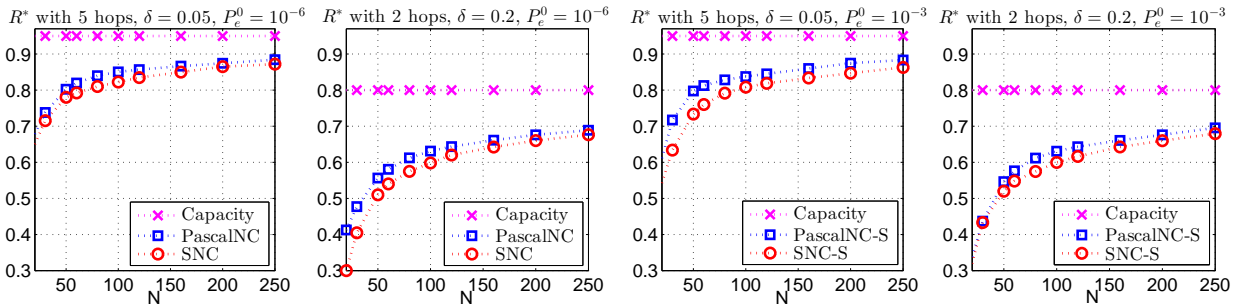,width=1.0\textwidth}
\end{center}
\caption{Optimal achievable rate at the destination, $R^{*}$, for different NC schemes over 2 and 5-hop line networks with $P_e^0=10^{-6}$ (subplot 1 \& 2) and $P_e^0=10^{-3}$ (subplot 3 \& 4). The PLR for the codes is averaged via 10000 simulation rounds.
\label{fig:THROUGHPUT_sim}}
\end{figure*}

\subsection{Optimal Achievable Rate}
\begin{figure*}[htbp]
\begin{centering}
\subfloat[Achievable rate				\label{fig:THROUGHPUT_Pe10e3}]				{\epsfig{file=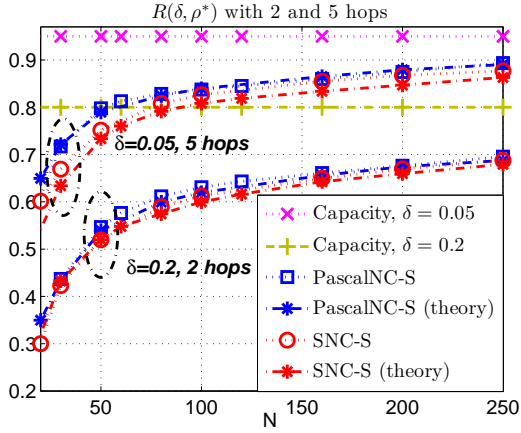,width=0.48\textwidth}}
\hfill
\subfloat[Packet delay					\label{fig:DELAY}]										{\epsfig{file=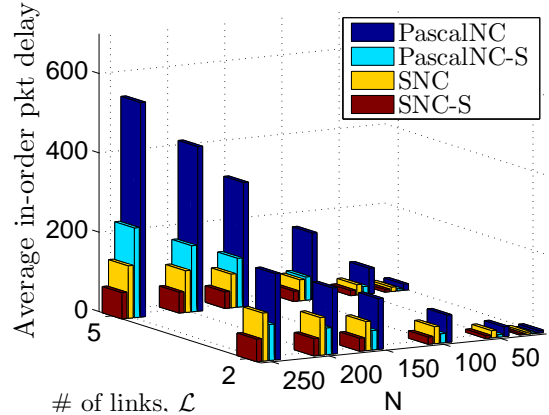,width=0.48\textwidth}}
\par\end{centering}
\caption{(a) Optimal achievable rate at the destination, $R^{*}$, for PascalNC-S and SNC-S over 2 and 5-hop line networks, $P_e^0=10^{-3}$. The PLR is averaged via simulations and also calculated using expressions: PascalNC-S (Eqs. (\ref{eq:PascalNC_S1})-(\ref{eq:PE_PascalNC_S})) and SNC-S (deduced from Eqs. (\ref{eq:PascalNC_S1})-(\ref{eq:PascalNC_S3})). (b) Average in-order packet delay at the destination node (in timeslots) for different NC schemes, $\delta=0.05$, $P_e^0=10^{-3}$.
\label{fig:THROUGHPUT_DELAY}}
\end{figure*}

We illustrate the resulting optimal achievable rate, $R^{*}$, from the $\rho^*$ obtained with Algo. \ref{Algo:SEARCHING} in Figs. \ref{fig:THROUGHPUT_sim} and \ref{fig:THROUGHPUT_Pe10e3} for Pascal and random codes. For $P_e^0=10^{-6}$, the logical choice is codes with no scheduling while for $P_e^0=10^{-3}$, we choose codes with scheduling to better control the delay.
For both theory and simulation, we observe that for $N\le 100$, $R^{*}$ follows an exponential behavior. Hence, our optimized coding rates allow the NC application designer to best tune the coding rate to avoid unnecessary decoding complexity (which increases cubically with generation size \cite{Li.2003}). Also, for the same $\mathcal{L}$ and $\delta$, $R^{*}$ with Pascal codes outperform random codes due to decoding and encoding at intermediate nodes.

Furthermore, Fig. \ref{fig:THROUGHPUT_Pe10e3} also shows that there is a gap between simulation curves and theory ones. The reason is that when calculating the theoretical optimal achievable rate at the destination, we use the theoretical upper-bound of the PLR for PascalNC-S (Eq. (\ref{eq:PE_PascalNC_S})) and for SNC-S (deduced from Eqs. (\ref{eq:PascalNC_S1})-(\ref{eq:PascalNC_S3})). Therefore, the resulting theoretical $R^{*}$ w.r.t. a specific network code would be lower than the corresponding simulation results. The gap increases with the number of links between the source-destination pair in the line path $\mathcal{G}$. 
Even if our theoretical results are conservative, we still obtain better performance compared to other implementations.

\subsection{Average In-order Packet Delay}
Let $\tau_j$ denote the time a packet $j$ is delivered in-order to the destination. For the set $\mathcal{S^*}$ packets delivered successfully, average in-order packet delay is computed as $T(\delta,\rho)=\frac{\sum^{|\mathcal{S^*}|}_{j=1}\tau_j}{|\mathcal{S^*}|}$.
Fig. \ref{fig:DELAY} shows average in-order packet delay w.r.t. $\rho^*$ for different NC schemes via simulations. We observe that the linear increase of the packet delay with $N$ and $\mathcal{L}$ is in accordance with the literature \cite{Dikaliotis.2009}. The figure shows that there is an achievable rate-delay tradeoff as Pascal codes obtain higher achievable rate with increased delay.

\subsection{Example Use Case}
We study a device-to-device (D2D) network \cite{WANG.2017} for emergency scenarios where a source transmits video packets to a destination over a line network with $\delta=0.1$ representing 802.11 links. Each node sends a packet in a timeslot with flow bit-rate of $2.5$ Mbps. The slot duration is $\frac{M*8}{2500} = 4.8$ ms. We assume a practical value of block-length e.g. $N=50$. For Pascal codes, we consider the offline optimized choice of Pascal matrix columns to generate the coefficient matrix. Hence, it needs no overhead to signal columns to the receivers.

Simulation results indicate that PascalNC-S can provide significant improvements on the achievable rate compared to the random codes. For example, for $\mathcal{L}=6$ and $\mathcal{L}=10$, the achievable rate gain with PascalNC-S can reach some $5.9\%$ and $13.2\%$ while the delay increase is $10.8\%$ and $58.7\%$ compared to SNC, respectively. 
Accordingly, coding overhead reduces from $0.4\%$ and $0.67\%$ to zero when random seeds are used and from $2.1\%$ and $1.9\%$ to zero when random seeds are not used.
Hence, the selection of an appropriate code should take into account both the gain in achievable rate and coding overhead, and delay constraints. 
Table \ref{tab:USE_CASES} shows an example for the selection of network codes such that achievable rate is maximized given target PLR and delay constraints \cite{Chen.2004}.
\begin{table}
\centering
\begin{tabular}{|c|c|c|c|c|} 
\cline{3-5}
\multicolumn{2}{c|}{\multirow{2}{*}{\begin{tabular}[c]{@{}l@{}}\textbf{Network code} \\\textbf{selected} \end{tabular}}} & \multicolumn{3}{c|}{\textbf{\# of links btw the source \& destination}}  \\ 
\cline{3-5}
\multicolumn{2}{c|}{}                           																												& \textbf{2-4} 		& \textbf{5-10} 		& \textbf{$>$ 10}           \\ 
\hline
\multirow{3}{*}{\begin{tabular}[c]{@{}l@{}}\textbf{Delay} \\\textbf{constraint} \\\textbf{(milisec)}\end{tabular}} & \textbf{100}           & PascalNC-S							& SNC-S  					&   					                  \\ 
\cline{2-5}
																																												& \textbf{200}           & PascalNC-S			& SNC			& SNC            	      					\\ 
\cline{2-5}
																																												& \textbf{300}          & PascalNC-S  			& PascalNC-S  		& SNC      \\
\hline
\end{tabular}
\caption{Example of optimal selection of network codes for the example use case w.r.t. different number of links and delay constraints for video services.
\label{tab:USE_CASES}}
\end{table}

\section{Conclusions \label{Sec:CONCLUSIONS}}
In this letter, we propose Pascal matrix-based structured codes and obtain their theoretical PLR. We then compare their finite-length performance with random codes by optimizing the coding rate for finite coding block sizes and assuming line networks. Our results reveal that our algebraic codes are a better choice than random codes for some practical applications.

\bibliographystyle{elsarticle-num}
\bibliography{DETERMINISTIC}

\end{document}